\documentclass[aps,prl,superscriptaddress,twocolumn,no-footinbib,showpacs,floatfix]{revtex4-1}

\usepackage{graphicx}
\usepackage{dcolumn}
\usepackage{bm}
\usepackage{color}
\usepackage[colorlinks=true,urlcolor=blue,citecolor=blue,linkcolor = blue]{hyperref}
\usepackage{amssymb,ulem,amsmath}

\newcommand{\qql}{\textquotedblleft}
\newcommand{\qqr}{\textquotedblright}
\newcommand{\vc}[1]{\bm{\mathrm{#1}}}

\begin{document}

\title{Edge binding of sine-Gordon solitons in spin-orbit coupled Bose-Einstein condensates}
\author{Sebastiano Peotta}
\email{speotta@physics.ucsd.edu}
\affiliation{Department of Physics, University of California, San Diego, La Jolla, CA 92093, USA}
\author{Francisco Mireles}
\email{fmireles@cnyn.unam.mx}
\affiliation{Department of Physics, University of California, San Diego, La Jolla, CA 92093, USA}
\affiliation{Centro de Nanociencias y Nanotecnolog{\'i}a, Universidad Nacional Aut\'onoma de M\'exico, Apdo. Postal 14, 22800 Ensenada, Baja California, M\'exico}
\author{Massimiliano Di Ventra}
\email{diventra@physics.ucsd.edu}
\affiliation{Department of Physics, University of California, San Diego, La Jolla, CA 92093, USA}

\begin{abstract}
In recent experiments with ultracold gases a Raman coupling scheme is used to produce both spin-orbit (SO) and Zeeman-type couplings [Y.-J. Lin \textit{et al.}, \href{http://www.nature.com/nature/journal/v471/n7336/abs/nature09887.html}{Nature \textbf{471}, 83 (2011)}]. Their competition drives a phase transition to a magnetized state with broken $Z_2$ symmetry. Using a hydrodynamic approach we study a confined binary condensate subject to both SO and Zeeman-type couplings.
We find that in the limit of small healing length  and in the phase with unbroken symmetry, the boundary magnetization profile has an analytical solution in the form of a sine-Gordon soliton. The soliton is bound to the edge of the system by the nontrivial boundary condition resulting from the combined effect of the SO coupling and the drop in the particle density. The same boundary condition is important in the magnetized phase as well, where we characterize numerically the boundary spin structure. We further discuss how the nontrivial magnetization structure affects the density profile near the boundary, yet  another prediction that can be tested in current experiments of spin-orbit coupled condensates.
\end{abstract}

\maketitle

\textit{Introduction} --- Ultracold atomic gases allow a relatively easy way to study several phenomena that are difficult to probe in solid-state
systems~\cite{Bloch:2008,Nori:2014}. Of particular note are those phenomena associated with spin-orbit (SO) coupling, which has been artificially engineered in a Bose-Einstein condensate (BECs) in Ref.~\onlinecite{Lin:2011} using a Raman laser scheme~\cite{Goldman:2013}. This opens
up the possibility of exploring the physics of bosons with SO coupling that was previously inaccessible.

In addition to the SO coupling, however, present laser schemes introduce also a finite Zeeman-type coupling~\cite{Lin:2011, Wang:2012,Cheuk:2012,Zhang:2012a,Goldman:2013}. While the SO coupling strength can be tuned only by changing the angle between the two Raman beams, the Zeeman-type field (ZF) is controlled by the intensity of the laser  beams. It follows that the two couplings cannot be separated. Their interplay manifests in a number of interesting effects, such as dipole oscillations~\cite{Zhang:2012a,Li:2012a}, Zitterbewegung~\cite{Zhang:2012b,Qu:2013,LeBlanc:2013}, pairing in Fermi gases~\cite{Fu:2013}, phase transitions~\cite{Stanescu:2008,Wang:2010,Ho:2011,Li:2012b,Ozawa:2013a}, excitation spectrum~\cite{Martone:2012,Li:2013}, break-down of Galilean invariance~\cite{Ozawa:2013b} and solitary waves~\cite{Merkl:2010,Zulicke:2012,Achilleos:2013}, to mention a few.
More details can be found in up-to-date reviews~\cite{soc_reviews}. 

Interestingly, in the solid state, it was already noticed that strong SO coupling and Zeeman gap (opened by an external magnetic field) are important ingredients for realizing exotic localized boundary states in nanowires~\cite{Lutchyn:2010,Oreg:2010}. However, the same type of analysis has not yet been considered in the context of cold gases (see however Ref.~\onlinecite{Baym:2012}).

In this Letter we focus on the boundary structure of BECs subject to both SO coupling and ZF. By means of a hydrodynamic approach~\cite{Lamacraft:2008,Refael:2009,Han:2012} we first recover the $Z_2$ symmetry breaking transition driven by the competition between SO and ZF~\cite{Lin:2011,Zhang:2013}. Next, we investigate the peculiar magnetization structure at the boundaries of a condensate confined by a box potential as realized in Ref.~\onlinecite{Hadzibabic:2013,Hadzibabic:2014}~\cite{foot4}.
In the limit where the healing length $\xi$ is small compared to the typical length scale $\ell_s$ of the spin degrees of freedom (defined below) the magnetization is controlled by a modified $O(3)$ sigma model with a boundary condition of the Neumann type induced by the sharp density drop at the potential step. Below the critical value of the SO coupling strength the boundary magnetization has the simple analytical form of a sine-Gordon soliton, which is bound to the edge due to the boundary condition above. The numerical solution of the Gross-Pitaevskii equation confirms that this solution is exact in the limit $\xi/\ell_s \to 0$. In the broken symmetry phase the boundary structure can be analysed only numerically taking into account the same boundary condition. Finally, we study how the  magnetization structure at the edges affects the density profile to the next leading order in $\xi/\ell_s$. All of our predictions are easily accessible experimentally in current cold gases schemes with artificial SO coupling~\cite{Lin:2011} and confined box-shaped potentials~\cite{Hadzibabic:2013,Hadzibabic:2014}.

\textit{Model system and hydrodynamic equations} ---  We use the standard single-particle Hamiltonian with SO coupling induced by Raman lasers~\cite{Lin:2011,soc_reviews}
\begin{equation}\label{eq:single-particle}
H_{\text{s.p.}} = -\frac{\hbar^2}{2m}\nabla_{\vc{r}}^2 -i\hbar\gamma\sigma_z\partial_x - \frac{\hbar \Omega}{2}\sigma_x + V(\vc{r})\,.
\end{equation}
The SO coupling strength $\gamma$ is given by $\gamma=\hbar k_L/m$ with $k_L$ the modulus of the difference of the Raman lasers wavevectors~\cite{Lin:2011,Goldman:2013}. The Raman coupling $\Omega$ introduces the Zeeman term $\frac{1}{2}\hbar\Omega\sigma_x$, $\sigma_{x,z}$ are the Pauli matrices, and $V(\vc{r})$ is the external potential.
\begin{figure*}
\includegraphics{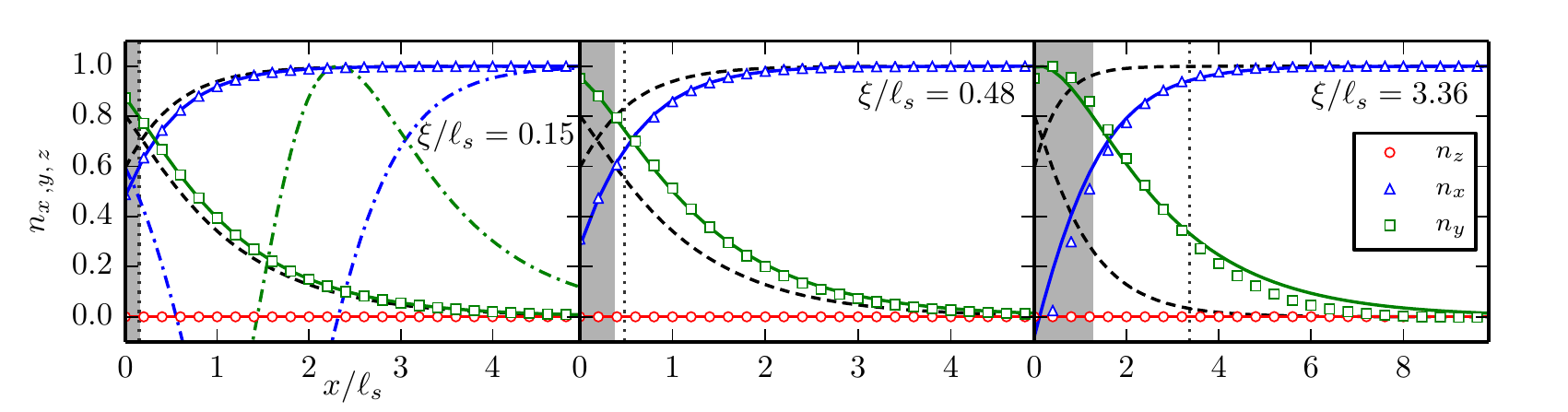}
\caption{\label{fig:one}(Color online) The magnetization obtained by imaginary-time evolution of the Gross-Pitaveskii equation~(\ref{eq:GP_eq}) is compared to the analytical formulas~(\ref{eq:sol1})-(\ref{eq:sol2}) for $\xi/\ell_s= 0.15$ (left panel), $\xi/\ell_s= 0.48$ (center panel) and $\xi/\ell_s= 3.36$ (right panel), while $c = 0.9$ in all panels. The triangles, squares and circles are respectively the $x,\,y,\,z$ components of the magnetization $\vc{n}(x)$ obtained numerically, while the solid lines are fits of the function $\phi_-(x-\xi^*,\ell_s^*)$ [Eqs.~(\ref{eq:sol1})-(\ref{eq:sol2}) and $x_0 = x_{0,+}$] with fit parameters $\xi^*$ and $\ell_s^*$: $\xi^*/\xi = 0.92$, $\ell_s^*/\ell_s = 1.01$ (left panel); $\xi^*/\xi = 0.75$, $\ell_s^*/\ell_s = 1.08$ (center); $\xi^*/\xi = 0.37$, $\ell_s^*/\ell_s = 2.05$ (right). The black dashed lines are the components $n_x$ and $n_y$ obtained from Eq.~(\ref{eq:sol1})-(\ref{eq:sol2}) without fitting ($\xi^*=0$ and $\ell_s^* = \ell_s$). The healing length $\xi$ is indicated by a vertical dotted line, while the grey area on the left of the panels has a width equal to $\xi^*$. The dashed-dotted lines in the left panel represent the solution corresponding to $x_{0,-}$ in Eq.~(\ref{eq:sol2}).}
\end{figure*}
Interactions are accounted for at the mean field level by the Gross-Pitaevskii equation (GPE)~\cite{Goldman:2013,soc_reviews}
\begin{equation}\label{eq:GP_eq}
i\hbar\partial_t \Psi = H_{\text{s.p.}}\Psi + g|\Psi|^2\Psi\,,
\end{equation}
which is a dynamical equation for the two-component (pseudospin-1/2) complex order parameter $\Psi(\vc{r}) = (\Psi_\uparrow(\vc{r}),\Psi_\downarrow(\vc{r}))$, and $H_{\text{s.p.}}$ is the Hamiltonian~(\ref{eq:single-particle}). The interaction term is isotropic in spin space, a good approximation for experiments~\cite{Lin:2011,Khamehchi:2014}, and is parametrized by the coupling constant $g$.


The condensed gas can be equivalently described by a complete set of gauge invariant quantities such as particle density $\rho(\vc{r})$, velocity field $\vc{v}(\vc{r})$~\cite{foot1} and magnetization vector $\vc{n}(\vc{r})\,\,(\vc{n}\cdot\vc{n} = 1)$.
This is the so-called hydrodynamic approach which has been introduced in Ref.~\cite{Lamacraft:2008,Refael:2009} for a spinor condensate and in Ref.~\cite{Han:2012} for the case of an isotropic Dresselhaus SO term. The case of one dimensional SO as in Eq.~(\ref{eq:single-particle}), is a straightforward  generalization~\cite{foot2},  thus only the final result is reported in the following.
After parametrizing the magnetization as $\vc{n}(\vc{r})=(\sin\theta(\vc{r})\cos\phi(\vc{r}),\sin\theta(\vc{r})\sin\phi(\vc{r}),\cos\theta(\vc{r}))$, where $\theta$ and $\phi$ are the polar an azimuthal angles, respectively, the energy functional of the condensate \textit{at rest} --- i.e. with $\vc{v} =0$, $\partial_t \rho = 0$ and $\partial_t \vc{n} = 0$ --- reads
\begin{gather}
\label{eq:erg_func_tot}
E = \int d^3x\left[\frac{g\rho^2}{2} + V\rho +\frac{(\hbar\nabla\sqrt{\rho})^2}{2m} + \rho H_{\sigma}\right]\,,\\
\label{eq:erg_func_spin}
H_{\sigma}=\frac{\hbar^2}{4m}\left[\frac{\left(\vc{\nabla}\theta\right)^2}{2}+\frac{\sin^2\theta}{2}\left(\vc{\nabla}\phi+\frac{c}{\ell_s}\hat{x}\right)^2 - \frac{\sin\theta\cos\phi}{\ell_s^2}\right].
\end{gather}
The spin length scale \, $\ell_s = \sqrt{\hbar/(2m\Omega)}$ and the reduced SO coupling strength $c = \gamma/\gamma_{\text{crit}}$  have been introduced, with $\gamma_{\text{crit}} = \sqrt{\hbar\Omega/2m}$ the SO coupling critical value.
$H_{\sigma}$ is the part of the energy functional that depends only on the magnetization and can be identified with the Hamitonian density of a $O(3)$ nonlinear sigma model~\cite{Rajaraman:1982} modified by SO coupling and ZF.
In a homogeneous system the minimum of energy is reached  for constant density $\rho(\vc{r}) = \bar\rho$ and $\phi=0$, while $\sin\theta = 1$ for $\gamma \leq \gamma_{\text{crit}}$ and $\sin\theta = 1/c^2$ for $\gamma > \gamma_{\text{crit}}$. In the latter case, $n_z$ has two possible values, $n_z = \pm\sqrt{1-1/c^{4}}$. Therefore the $Z_2$ symmetry of the Hamiltonian under the transformation $\sigma_z \to -\sigma_z$ and $x \to -x$ is broken. This corresponds precisely to the phase transition resulting from the competition of the ZF and the  SO coupling~\cite{Lin:2011,soc_reviews}. In the following we consider the cases $\gamma \leq \gamma_{\text{crit}}$ and $\gamma> \gamma_{\text{crit}}$ separately.

\textit{The $\gamma \leq \gamma_{\text{crit}}$ case} ---
When $\gamma \leq \gamma_{\text{crit}}$ we assume the condition $\sin\theta=1$ ($n_z = 0$) for a  inhomogeneous system as well  and consider a translational invariant system in the $y$ and $z$ directions.
The two Euler-Lagrange equations obtained by varying the energy functional~(\ref{eq:erg_func_tot})-(\ref{eq:erg_func_spin}) are
\begin{gather}
\label{eq:euler_lagrange_rho}
g\rho -\frac{\hbar^2\partial_x^2\sqrt{\rho}}{2m\sqrt{\rho}} + \frac{\hbar^2}{4m}\left[\frac{1}{2}\left(\partial_x\phi+\frac{c}{\ell_s}\right)^2-\frac{\cos\phi}{\ell_s^2}\right] = V\,,
\\
\label{eq:euler_lagrange_phi}
\frac{\hbar^2}{4m}
\left[\partial^2_x\phi + \frac{\partial_x\rho}{\rho}\left(\partial_x\phi+\frac{c}{\ell_s}\right)\right] = \frac{\hbar\Omega}{2}\sin\phi\,.
\end{gather}
For $c=0$ the last two terms on the left-hand side of Eq.~(\ref{eq:euler_lagrange_rho}) form together the sine-Gordon Hamiltonian density. Eq.~(\ref{eq:euler_lagrange_phi}) is precisely the static sine-Gordon equation with an additional term $\propto \partial_x\rho/\rho$ on the left hand side.


In the limit of very large interaction strength $g$, the density $\rho$ and the angle $\phi$ in Eqs.~(\ref{eq:euler_lagrange_rho})-(\ref{eq:euler_lagrange_phi}) decouple since the density is very stiff and slightly affected by a nonuniform magnetization. This can be shown by considering a hard-wall confining potential of the form $V(x\geq 0) = \mu$, $V(x<0) = +\infty$ at the left boundary. The solutions of Eq.~(\ref{eq:euler_lagrange_rho}), neglecting the $\phi$-dependent part, is
\begin{equation}\label{eq:density_boundary}
\rho_0(x) = \bar\rho \tanh^2\frac{x}{\xi}\qquad \text{for}\quad x\geq 0\,.
\end{equation}
The scale of density variation is the healing length $\xi = \sqrt{\hbar^2/(m \mu)}$ and $\bar\rho = \mu/g$ is the background density far from the boundary. The limit of $\partial_x\rho_0/\rho_0$ for $\xi \to 0$ ($g\to +\infty$) is a discontinuous function~\cite{foot5}
\begin{equation}
\lim_{\xi\to 0} \frac{\partial_x\rho_0}{\rho_0} =
\begin{cases}
+\infty, &x = 0\,, \\
0, &x >0\,.
\end{cases}
\end{equation}
Therefore, Eq.~(\ref{eq:euler_lagrange_phi}) reduces to a pure sine-Gordon equation $\partial_x^2\phi =\sin\phi/ \ell^2_{s}$ with the non-trivial boundary condition
\begin{equation} \label{eq:boundary_condition}
\left.\partial_x\phi\right|_{x=0} = -\frac{c}{\ell_s} = -\frac{2m\gamma}{\hbar}\,.
\end{equation}
The localized solutions of the sine-Gordon equation have the form~\cite{Rajaraman:1982,Manton:2004}
\begin{equation}\label{eq:sol1}
\phi_\pm(x,\ell_s) =  \pm4\arctan e^{(x+x_0)/\ell_s}\, ,
\end{equation}
and are called the soliton ($\phi_+$) and antisoliton ($\phi_-$), respectively.
The (anti-)soliton solution is parametrized by the coordinate $-x_0$ of  the \qql center of mass\qqr because of the translationally invariance of the sine-Gordon equation. Translational invariance is broken at the boundary and in fact the boundary condition~(\ref{eq:boundary_condition}) can be satisfied for $\gamma>0$ by the antisoliton solution ($\phi_-$) with center of mass coordinate of the form
\begin{equation}\label{eq:sol2}
x_{0, \pm} = \ell_s\ln\left(2 c^{-1}\pm\sqrt{4c^{-2}-1}\right)\,.
\end{equation}
The ground state solution for the \textit{left} boundary is the one corresponding to the plus sign in Eq.~(\ref{eq:sol2}) since it is continuously connected to the correct ground state in the small SO coupling limit. Moreover, it is easy to verify that this solution has lower energy with respect to the other one. The solution with minus sign in Eq.~(\ref{eq:sol2}) corresponds to the \textit{right} boundary. In the following we consider only the left boundary and we take $x_0 = x_{0,+}$.

\begin{figure}[t]
\includegraphics{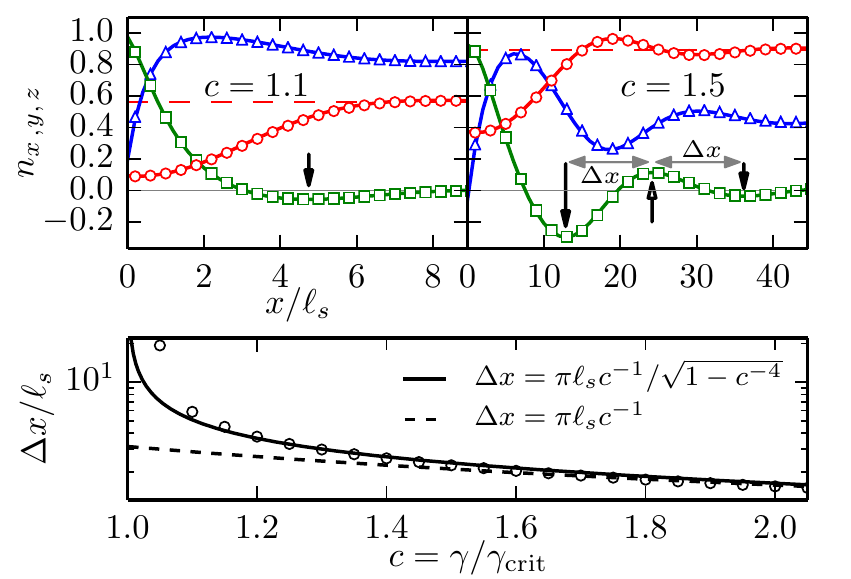}
\caption{\label{fig:two} (Color online) Upper panel: Components of the magnetization $n_x$ (triangles), $n_y$ (squares) and $n_z$ (circles) obtained by numerical solution of the GPE in the case $\gamma > \gamma_\text{crit}$, while the solid lines are the corresponding solutions that minimize the modified nonlinear sigma model~(\ref{eq:erg_func_spin}) with the boundary conditions $\partial_x \theta|_{x=0}=0$ and Eq.~(\ref{eq:boundary_condition}), shifted to the right by $\xi$ (without fitting). Two values of the SOC have been considered: $c = 1.1$ (left upper panel) and $c=1.5$ (right upper panel), while $\xi/\ell_s = 0.15$ in both panels. The dashed horizontal lines denote the non-zero asymptotic value of $n_z(x)$  ($\,n_z|_{x\to +\infty}= \sqrt{1-1/c^4}\,$), while the arrows correspond to the minima and maxima of $n_y(x)$. Lower panel: The circles are the distance $\Delta x$ between the first minimum and the neighbouring maximum of $n_y$ as function of $c$ above the critical point, while the solid (dashed) line is the analytic form of the spin-precession length calculated with (without) the ZF. The expression including the ZF is closer to the numerical data.}
\end{figure}

Eqs.~(\ref{eq:sol1})-(\ref{eq:sol2}) provide an analytical form for the magnetization profile at the left boundary of the system when $c < 1$. 
We provide below numerical and analytical arguments showing that this solution is in fact exact in the $\xi/\ell_s \to 0$ limit. Moreover we show numerically that it captures approximately the magnetization profile in the regime $\xi \lesssim \ell_s$. To this end, we consider the solution~(\ref{eq:sol1})-(\ref{eq:sol2}) as a fitting function $\phi_-(x-\xi^*,\ell_s^*)$ where the fitting parameters $\xi^*$ and $\ell_s^*$ allow for translation and rescaling, respectively.
This function is fitted to the numerical solution of the GPE~(\ref{eq:GP_eq})~\cite{foot3} for three different values of the ratio 
$\xi/\ell_s$, namely $\xi/\ell_s = 0.15,\,\,0.48,\,\,3.36$, while $c = \gamma/\gamma_{\text{crit}}=0.9$ is fixed, and the results are shown in Fig.~\ref{fig:one}. For $\xi/\ell_s \ll 1$ the agreement with the numerical data is excellent with $\xi^*$ and $\ell_s^*$ very close to $\xi$  and $\ell_s$, respectively. The fact that $\xi^* \sim \xi$ is expected since in the derivation of Eqs.~(\ref{eq:sol1})-(\ref{eq:sol2}) we neglected that the density drops to zero on a finite width $\xi$ at the boundary. As shown in Fig.~\ref{fig:one} the fitted solution is a very good approximation for $\xi \lesssim \ell_s$, while for $\xi > \ell_s$ some deviations are noticeable. With increasing ratio $\xi/\ell_s$ the fitted value $\xi^*$ tends to decrease while $\ell_s^*$ tends to increase. We have obtained similar results in the whole range $0 \leq c \leq 1$. Within numerical precision we conclude that the solution~(\ref{eq:sol1})-(\ref{eq:sol2}) is exact in the limit $\xi/\ell_s \to 0$ $(g \to +\infty)$ limit, and that for finite $g$ the assumption $n_z = 0$ holds. The same results apply for the right boundary with the only difference that $x_{0,+}\to x_{0,-}$ and consequently $n_y\to -n_y$.


\textit{The $\gamma>\gamma_\text{crit}$ case} ---
For $\gamma > \gamma_\text{crit}$ we are not able to provide an analytical solution in the limit $\xi/\ell_s \to 0$, but we can still minimize numerically the functional in Eq.~(\ref{eq:erg_func_spin}) with the proper boundary condition. One easily finds that the boundary condition $\partial_x \theta|_{x=0}=0$ needs to be enforced in addition to  Eq.~(\ref{eq:boundary_condition}). In the upper panel of Fig.~\ref{fig:two} we show the boundary structure of a condensate with SO coupling strength $c=1.1$ (left) and $c= 1.5$ (right). The solutions of the nonlinear sigma model are shown as lines in Fig.~\ref{fig:two} (upper panel). These have been shifted by $\xi$ and provide a significantly better agreement with the full solution of the GPE, as in the case $\gamma \leq \gamma_\text{crit}$.

New features for $\gamma>\gamma_\text{crit}$ are: i) $n_z(x)$ is non-uniform and increases from
$n_z(x=0) >0$ up to its asymptotic value on a length scale that increases when approaching the critical point; ii) oscillations in all the magnetization components appear.
We also found that the distance between the first minimum of $n_y$ and the neighbouring maximum is given, to a good approximation, by the spin-precession length~\cite{Datta:1990}, which is the wavelength corresponding to the minima of the dispersion of $H_\text{s.p.}$~(\ref{eq:single-particle}). The expression for the spin-precession length is provided in the legend of Fig.~\ref{fig:two} (lower panel).

\begin{figure}
\includegraphics{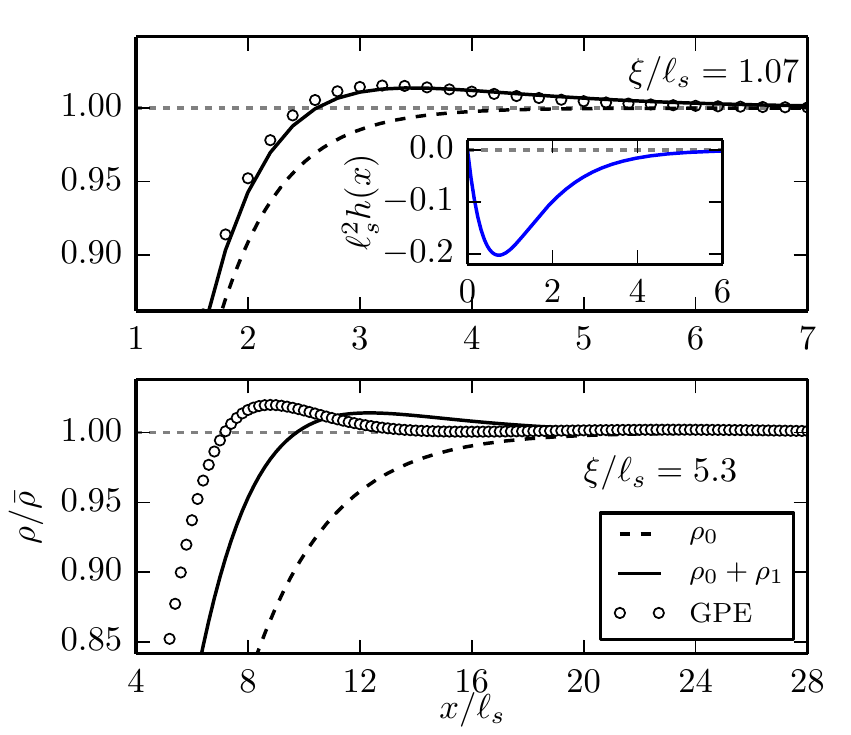}
\caption{\label{fig:three} (Color online) Density profiles obtained by means of the GPE (circles) compared to the zero order result $\rho_0$~(\ref{eq:density_boundary}) (dashed line) and to the first order one $\rho_0+\rho_1$~(\ref{eq:back-action}) (solid line). Two different values of the ratio $\xi/\ell_s$ are considered: $\xi/\ell_s = 1.07$ (upper panel) and $\xi/\ell_s= 5.3$ (lower panel). The reduced SO coupling strength is $c=0.9$ as in Fig.~\ref{fig:one}. The energy density $h(x)$ has been calculated with the function $\phi_-(x-\xi^*,\xi^*_s)$ fitted as in Fig.~\ref{fig:one} [fit parameters $\xi^*/\xi =0.68$, $\ell_s^*/\ell_s = 1.26$ (upper); $\xi^*/\xi = 0.27$, $\ell_s^*/\ell_s = 2.45$ (lower)]. In inset the energy density $h(x)$ in Eq.~(\ref{eq:soliton_energy}) is shown.}
\end{figure}

\textit{Density profile} --- In the previous discussion we have assumed the density profile at the boundary to be as in Eq.~(\ref{eq:density_boundary}), namely unaffected by the magnetization. We now provide an estimate of the back-action of the non-trivial magnetization structure at the boundary on the density profile, thereby rigorously justifying the result of Eqs.~(\ref{eq:density_boundary})-(\ref{eq:sol2}). The first order correction $\rho_1(x)$ to Eq.~(\ref{eq:density_boundary}) can be expressed as
\begin{equation}\label{eq:back-action}
\frac{\rho_1(x)}{\bar\rho} = -\xi^2\tanh\frac{x}{\xi}\int_0^{+\infty} dz\, G(x/\xi,z)\tanh(z)\, h(\xi z)\,.
\end{equation}
Here we use the dimensionless variable $z = x/\xi$ and $G(z,z')$ is the Green's function of the differential operator $\mathcal{L} = -\frac{d^2}{dz^2}+2(3\tanh^2 z-1)$, which is obtained by linearizing Eq.~(\ref{eq:euler_lagrange_rho}) around the equilibrium solution~(\ref{eq:density_boundary}). The Green's function has the form $G(z,z') = G(z',z) = y_1(z)y_2(z')$ for $z>z'$ with $y_i(z)$ satisfying $\mathcal{L}\,y_i(z) = 0$, $y_1 \sim e^{-2z}$ for $z\to +\infty$, $y_2(0) = 0$ and $\frac{dy_2}{dz}(0) = \frac{1}{y_1(0)}$.
The energy density $h(x) = \frac{4m}{\hbar^2}H_{\sigma}(x)$~(\ref{eq:erg_func_spin}) enters Eq.~(\ref{eq:back-action}) and for the sine-Gordon soliton~(\ref{eq:sol1}) it reads
(constant terms are reabsorbed in the chemical potential)
\begin{equation}\label{eq:soliton_energy}
 h(x) = \frac{4}{\ell_s^2\cosh\frac{x+x_0}{\ell_s}}\left(\frac{1}{\cosh\frac{x+x_0}{\ell_s}}-\frac{1}{\cosh\frac{x_0}{\ell_s}} \right)\,.
\end{equation}
Since the Green's function can be roughly approximated by the asymptotic form $G(z,z') = e^{-2|z-z'|}/2$, Eqs.~(\ref{eq:back-action})-(\ref{eq:soliton_energy}) show that the correction is first order in the expansion parameter $(\xi/\ell_s)^2 = 2\hbar\Omega/\mu$. 

In Fig.~\ref{fig:three} the possible approximations to the exact density profiles are compared.
The first order correction is already able to capture the characteristic non-monotonic behavior of the density near the boundary. The density bump visible in Fig.~\ref{fig:three} matches the energy density of the soliton which is negative and strongly localized (see inset of Fig.~\ref{fig:three}). Again, we find good agreement up to $\xi \lesssim \ell_s$, while the lower panel of Fig.~\ref{fig:three} shows that only a rough qualitative agreement can be obtained for weak interactions, even after fitting the soliton shape as in Fig.~\ref{fig:one}.

The ratio $\xi/\ell_s$ is thus rigorously established as the small parameter for the approximation leading to Eqs.~(\ref{eq:sol1})-(\ref{eq:sol2}). For $^{87}$Rb ($g/\hbar = 52.6\times 10^{-12}\,\mathrm{Hz}\,\mathrm{cm}^{3}$) $\xi$ can vary from $1\,\mu$m to $0.1\,\mu$m in the density range $\bar{\rho}=10^{13}$ to $10^{15}$ cm$^{-3}$, 
while $\ell_s$ is of the order of the Raman laser wavelength $2\pi/k_L$ ($804\,\mathrm{nm}$ in Ref.~\cite{Lin:2011}). Therefore all the values of $\xi/\ell_s$ shown in Fig.~\ref{fig:one}-\ref{fig:three} are realistic.

\textit{Summary} --- In summary, we predict that the boundary condition that stems from the abrupt change in density at the
edge (on a scale $\xi$) of a confined BEC has the effect of binding a sine-Gordon soliton. The fingerprint
of this soliton is a finite component of the magnetization along the axis orthogonal both to the Zeeman term axis and spin-orbit axis, and is a combined effect of both terms. Above the phase transition, the same boundary condition is equally important and produces qualitatively similar magnetization profiles, but with an added oscillation on the scale of the spin-precession length. This predictions, together with the characteristic shape of the particle density near the boundary, are well within reach of present experiments. Our work is also a starting point for investigating the behavior of the system under a time-dependent gauge field~\cite{Chih-Chun:2013,Peotta:2014}.



\begin{acknowledgments}
SP and MD acknowledge support from DOE under Grant No. DE-FG02-05ER46204. FM acknowledges the support of DGAPA-UNAM via PASPA for the fellowship provided to carry out a sabbatical leave at the Department of Physics, UCSD.
\end{acknowledgments}


\begin{thebibliography}{77}
\bibitem{foot4} A flat confining potential is ideal to highlight the physics we are interested in, but the same phenomena are present in the more common case of harmonic confinement. The simple analytical solution presented in the following holds strictly only for a flat potential with sharp confining walls.
\bibitem{foot1}
For a many-body quantum system the velocity is not an observable by itself, since it is the ratio of two gauge invariant observable such as current and density. However when a mean-field equation is used, such as the Gross-Pitaevskii equation~(\ref{eq:GP_eq}),  there is no ambiguity.
\end{thebibliography}
\end{document}